\documentclass[12pt]{article}

\include{amssym}
\def\mathbf{\vec}

\newcommand{\njl}{\mathrm{NJL}}
\newcommand{\dt}{\mathrm{det}}
\newcommand{\inter}{\mathrm{int}}
\newcommand{\ud}{\mathrm{d}}
\newcommand{\st}{\mathrm{st}}
\newcommand{\symb}{\mathrm{SB}}
\newcommand{\e}{\mathrm{E}}
\newcommand{\tad}{\mathrm{tad}}
\newcommand{\kin}{\mathrm{kin}}
\newcommand{\mass}{\mathrm{mass}}
\newcommand{\ren}{\mathrm{R}}
\newcommand{\p}{\mathrm{p}}
\newcommand{\s}{\mathrm{s}}
\newcommand{\ns}{\mathrm{ns}}
\newcommand{\id}{\mathrm{id}}

\begin{document}

\centerline{\Large\bf Long distance expansion for the NJL model}
\vspace{0.5cm}

\centerline{\Large\bf with $SU(3)$ and $U_A(1)$ breaking}
\vspace{1cm}

\centerline{\large A.A. Osipov\footnote{Joint Institute for Nuclear Research, 
                                 Laboratory of Nuclear Problems, 
                                 141980 Dubna, Moscow Region, Russia.}, 
            H. Hansen, B. Hiller}
\vspace{0.1cm}

\centerline{\small\it Centro de F\'{\i}sica Te\'{o}rica, Departamento de
         F\'{\i}sica da Universidade de Coimbra,}
\centerline{\small\it 3004-516 Coimbra, Portugal} 
\vspace{1cm}

\centerline{{\bf Abstract}}
\vspace{.5cm}
This work is a follow up of recent investigations, where we study the 
implications of a generalized heat kernel expansion, constructed to 
incorporate non-perturbatively the effects of a non-commutative  
quark mass matrix in a fully covariant way at each order of the expansion. 
As underlying Lagrangian we use the Nambu -- Jona-Lasinio model of QCD, 
with $SU_f(3)$ and $U_A(1)$ breaking, the latter generated by the 't Hooft 
flavour determinant interaction. The associated bosonized Lagrangian is 
derived in leading stationary phase approximation (SPA) and up to second 
order in the generalized heat kernel expansion. Its symmetry breaking 
pattern is shown to have a complex structure, involving all powers of 
the mesonic fields allowed by symmetry. The considered Lagrangian yields 
a reliable playground for the study of the implications of symmetry and 
vacuum structure on the mesonic spectra, which we evaluate for the scalar 
and pseudoscalar meson nonets and compare with other approaches and 
experiment. 

\newpage

\section*{\normalsize 1. Introduction}

The heat kernel expansion \cite{Vassilevich:2003} is known as a 
useful and effective tool to study the properties of low-energy QCD 
\cite{Ebert:1986,Espriu:1990,Bijnens:1993}. Depending on the physical 
problem, it can be used either in the form of a derivative expansion 
\cite{Chan:1986}, or as an inverse mass expansion \cite{Min:1982}. 
Based on the powerful method of Schwinger -- DeWitt 
\cite{Schwinger:1951}, it allows for calculations of effective meson 
Lagrangians directly in coordinate space by integrating out the 
quadratic fluctuations of quark fields in presence of a background
of classical mesonic fields. The result is cast as an asymptotic 
expansion of the effective action in powers of proper time with 
Seeley -- DeWitt coefficients $a_n$, which accumulate the whole 
dependence on the background fields. The remarkable property of the 
method is that each order of the expansion is fully gauge and chiral 
covariant. 

In the case of massive quantum fields with a degenerate mass matrix 
$M=\mbox{diag}(m,m,...)$, it is not difficult to derive from the proper 
time expansion an expansion in inverse powers of $m^2$, since the mass 
dependence is easily factorized and a subsequent integration over the 
proper time leads to the desired result. The resulting asymptotic 
coefficients remain unchanged. 

If the mass matrix is however non degenerate $M=\mbox{diag}(m_1,m_2,...)$ 
its total factorization is impossible because of the non-commutativity 
of the matrix $M$ with the rest of the elliptic operator. It has been shown 
recently \cite{Osipov1:2001,Osipov2:2001} that masses can be 
redistributed among the mass-dependent factors by performing resummations 
in the series. This leads to new covariant asymptotic coefficients. 
The algorithm for the resummations was derived and the generalized 
heat kernel coefficients $b_n$ for the $SU_f(2)$ \cite{Osipov1:2001} 
and $SU_f(3)$ \cite{Osipov2:2001} flavour cases were obtained. 
In \cite{Salcedo:2001} the relation of the new coefficients with the 
standard ones has been clarified.

Given the success in the mathematical formulation of the problem, it 
is now a natural step to apply the new asymptotic expansion in 
the construction of effective chiral Lagrangians. This expansion
provides a reasonable approximation to the physics of massive
and heavy quantum fields with a non-degenerate mass matrix. 
This is the case, for instance, of low energy QCD. Here a light 
current quark mass matrix which is non-degenerate is replaced by a 
non-degenerate mass matrix of heavy constituent quarks through
the non-perturbative mechanism of spontaneous breakdown of chiral 
symmetry. This area of physics opens a window where our generalized 
heat kernel expansion can be applied.        

Several different approaches based on the standard heat kernel series 
have already been used to study the above mentioned task 
\cite{Ebert:1986,Espriu:1990,Bijnens:1993}. The main difference 
between them is hidden in the definition of the vacuum state. 
The generalized heat kernel expansion also leads to its own 
prescription for the vacuum. It is clear that the closer one is 
to the physical vacuum, the more realistic the description of the 
spectrum of the mesonic exitations will be. From this point of view 
we hope that our method is a useful tool for the accurate 
description of the hadronic vacuum state at low energies.   

In the present work we shall choose a well-known quark model 
\cite{Nambu:1961} to describe the formation of the hadronic 
vacuum and its mesonic exitations. It is an effective microscopic low 
energy Lagrangian combining the $U_L(3)\times U_R(3)$ chiral four-quark 
Nambu -- Jona-Lasinio (NJL) interactions together with the 't 
Hooft determinantal six-quark flavour-mixing interaction, responsible 
for $U_A(1)$ breaking \cite{Hooft:1976}. By including a mass term for 
the light $u,d$ and strange $s$ quarks one can explicitly break the 
remaining $SU_L(3)\times SU_R(3)$ chiral symmetry to the $SU_f(3)$ 
flavour group or its subgroups. This Lagrangian has been previously 
used in \cite{Veronique:1988,Reinhardt:1988} to calculate the low 
lying meson mass spectrum at leading order\footnote{An early approach
but without 't Hooft term can be found in \cite{Volkov:1984}}. 
In the recent works \cite{Osipov7:2002} we have analyzed the 
quasi-classical corrections stemming from the 't Hooft interaction, 
and presented a fully analytical solution for the 
bosonized Lagrangian and effective potential. We will use here these 
results. They represent a necessary step for the extension to a
larger group of the earlier applications of the method in the 
$SU(2)\times SU(2)$ NJL model \cite{Osipov5:2001,Osipov6:2001}. 

The paper is organized as follows. In Section 2 we introduce the model 
and present the main results of \cite{Osipov7:2002} needed for the 
present work. To summarize, these results are the following. Using path
integral methods, the bosonization of the fermionic Lagrangian 
which involves the six quark interaction requires the introduction of 
two sets of bosonic auxiliary fields, each of the scalar and 
pseudoscalar type, say $(s,p)$ and $(\sigma,\phi )$. Then the 
integration over the fermionic fields can be cast in quadratic form, 
which can be done exactly. The remaining integrations are over one of 
the sets of auxiliary bosonic variables, $(s,p)$ which are done in the
stationary phase approximation. The solutions to the stationary path 
integral equations can be expressed as an infinite series in powers of 
the bosonic scalar and pseudoscalar fields $(\sigma,\phi )$, with 
coefficients that are known at all orders. In particular also the 
symmetry breaking piece of the bosonized Lagrangian contains an 
infinite number of terms involving powers of $(\sigma,\phi )$, and are 
a consequence of the flavour determinantal interaction. The piece of 
the bosonized Lagrangian which comes from the integration over the 
fermionic degrees of freedom will be dealt with our generalized heat 
kernel technique in Section 3. In Section 4 we show how to deal with 
the gap equations combined with the requirement of covariance of the 
generalized Seeley -- DeWitt coefficients and the symmetry breaking 
pattern of the original Lagrangian which must be not altered. We 
derive the expressions for the masses of the pseudoscalar and scalars 
in Section 5. In Section 6 we present numerical results and conclude 
with a summary and outlook.


\section*{\normalsize 2. The model}

To model low energy QCD, we use the global $U_L(3)\times U_R(3)$ chiral 
symmetric four-quark interaction of the NJL-type model
\begin{equation}
\label{L4q}
   {\cal L}_{\njl}=\frac{G}{2}\left[
            (\bar{q}\lambda_aq)^2+
            (\bar{q}i\gamma_5\lambda_aq)^2\right],
\end{equation}
where $\lambda_a,\ a=0,1,\ldots ,8$ are the standard Gell-Mann matrices 
acting in flavour space and normalized by the condition $\mbox{tr}(
\lambda_a\lambda_b)=2\delta_{ab}$, 
combined with the 't Hooft six-quark flavor determinantal 
interaction \cite{Hooft:1976}
\begin{equation}
\label{Ldet}
  {\cal L}_\dt = \kappa (\det\bar{q}P_Lq
                         +\det\bar{q}P_Rq),
\end{equation}
where the matrices $P_{L,R}=(1\mp\gamma_5)/2$ are projectors on the
left- and right-handed quarks. The 
Lagrangian ${\cal L}_{\dt}$ lifts the unwanted $U_A(1)$ 
symmetry of ${\cal L}_\njl$ for massless quarks, as required by 
the $U_A(1)$ Adler-Bell-Jackiw anomaly of the $SU_f(3)$ singlet 
axial current $\bar{q}\gamma_\mu\gamma_5q$ in QCD. The total 
fermionic Lagrangian reads
\begin{equation}
\label{totlag}
  {\cal L}=\bar{q}(i\gamma^\mu\partial_\mu -\hat{m})q
          +{\cal L}_{\inter},
\end{equation}
with the interaction Lagrangian
\begin{equation}
\label{L_int}
  {\cal L}_{\inter}={\cal L}_{\njl}+{\cal L}_{\dt}.
\end{equation}
The quark fields have color $(N_c=3)$ and flavor $(N_f=3)$ indices
which range over the set $i=1,2,3$. The current quark mass, $\hat{m}$, is a 
diagonal matrix with elements $\textrm{diag}(\hat{m}_u, \hat{m}_d,
\hat{m}_s)$, which explicitly breaks the global chiral $SU_L(3)\times 
SU_R(3)$ symmetry of the Lagrangian.
 
This approach contains several commonly used simplifications which 
can be excluded in a more elaborate consideration. 
Let us comment first on the four-point interaction (\ref{L4q}). The
most general form of this vertex, based on phenomenological arguments,
needs only to be compatible with the symmetry group of low energy QCD 
and can be chosen to be invariant under the $SU(3)_c\times SU_L(3)\times 
SU_R(3)\times U_V(1)\times U_A(1)$ group. The six-point 
interaction (\ref{Ldet}) corresponds to the $N_c\rightarrow \infty$ 
limiting case and is modified by the tensor term at next to the
leading $1/N_c$ order as it follows from the instanton dynamics 
\cite{Diakonov:1992}. We also assume that all interactions between quarks 
are taken in the long wavelength limit (low momenta) where they are
effectively local. The explicit chiral symmetry breaking term, 
$\bar q\hat mq$, is standard for QCD. There are some doubts  
in the literature regarding this structure in the context
of the NJL Lagrangian \cite{Carter:1999}. Further discussion
of this point based on an instanton approach to the QCD vacuum can be found 
in \cite{Lee:1979}.       
 
In order to access the natural degrees of freedom of low-energy QCD in
the mesonic sector, we proceed to bosonize the fermionic Lagrangian, by 
introducing in the vacuum persistence amplitude 
\begin{equation}
\label{genf1}
   Z=\int {\cal D}q{\cal D}\bar{q}\exp\left(i\int\! \ud^4 x 
          {\cal L}\right)
\end{equation}  
the functional unity \cite{Reinhardt:1988} 
\begin{eqnarray}
\label{1}
   1\!\!\!\!\!\!\!\!
   &&=\int \prod_a {\cal D}s_a{\cal D}p_a\delta (s_a-\bar{q}\lambda_aq)
      \delta (p_a-\bar{q}i\gamma_5\lambda_aq)
      \nonumber \\
   &&=\int \prod_a {\cal D}s_a {\cal D}p_a
     {\cal D}\sigma_a {\cal D}\phi_a  \nonumber \\
   &&\ \ \  
     \times\exp\left\{i\int\! \ud^4 x[\sigma_a(s_a-\bar{q}\lambda_aq)+
     \phi_a(p_a-\bar{q}i\gamma_5\lambda_aq)]\right\},
\end{eqnarray}
thus obtaining
\begin{eqnarray}
\label{genf3}
   Z\!\!\!\!\!\!\!\!
   &&=\int \prod_a{\cal D}\sigma_a{\cal D}\phi_a
          {\cal D}q{\cal D}\bar{q}
          \exp\left(i\int\!\ud^4 x{\cal L}_q(\bar{q},q,\sigma ,\phi )
          \right)\nonumber \\
   &&\ \ \ \times\int\prod_a{\cal D}s_a{\cal D}p_a
       \exp\left(i\int\!\ud^4 x{\cal L}_r(\sigma ,\phi ,s,p)\right),
\end{eqnarray}
where
\begin{eqnarray}
\label{lagr2}
   {\cal L}_q\!\!\!\!\!\!\!\!
   &&=\bar{q}(i\gamma^\mu\partial_\mu -\sigma 
              -i\gamma_5\phi )q, \\
\label{lagr3}
   {\cal L}_r\!\!\!\!\!\!\!\!
   &&=\frac{G}{2}\left[(s_a)^2+(p_a)^2\right]
            +s_a(\sigma_a -\hat{m}_a)+p_a\phi_a \nonumber \\
   &&\ \ \  +\frac{\kappa}{32}A_{abc}s_a\left(s_bs_c-3p_bp_c
            \right).
\end{eqnarray}
and where the totally symmetric constants $A_{abc}$ are related to the 
flavour determinant, and equal to
\begin{eqnarray}
\label{A}
   A_{abc}\!\!\!\!\!\!\!\!
   &&=\frac{1}{3!}\epsilon_{ijk}\epsilon_{mnl}(\lambda_a)_{im}
             (\lambda_b)_{jn}(\lambda_c)_{kl} \nonumber \\
   &&=\frac{2}{3}d_{abc} +
      \sqrt{\frac{2}{3}} \Big(
      3\delta_{a0}\delta_{b0}\delta_{c0}
      -\delta_{a0}\delta_{bc}
      -\delta_{b0}\delta_{ac}
      -\delta_{c0}\delta_{ab}\Big).
\end{eqnarray}
We use the standard definitions for antisymmetric $f_{abc}$ and
symmetric $d_{abc}$ structure constants of $U(3)$ flavour symmetry. 
One can find, for instance, the following useful relations
\begin{eqnarray}
   &&f_{eac}A_{bfc}+f_{ebc}A_{fac}+f_{efc}A_{abc}=0, \nonumber \\ 
   &&d_{eac}A_{bfc}+d_{ebc}A_{fac}+d_{efc}A_{abc} 
     = \sqrt{6}\delta_{e0}A_{abf}.
\end{eqnarray}

Here and throughout the paper we use $\sigma=\sigma_a\lambda_a$, and
so on for all auxiliary fields, $\phi ,\ s,\ p$, and use the following 
representation  of the scalar and pseudoscalar fields 
\begin{equation}
\label{fields}
  \frac{  \lambda_a\sigma_{a}}{\sqrt{2}}=\left(
          \begin{array}{ccc}
          \frac{\displaystyle \sigma_u}{\displaystyle \sqrt{2}}& 
          a_0^+ & K^{*+}_0 \\
          a_0^-& 
          \frac{\displaystyle \sigma_d}{\displaystyle \sqrt{2}}
          &K^{*0}_0  \\
          K^{*-}_0 & \bar{K}^{*0}_0 &
          \frac{\displaystyle \sigma_s}{\displaystyle \sqrt{2}}
          \end{array} \right),\qquad
  \frac{  \lambda_a\phi_{a}}{\sqrt{2}}=\left(
          \begin{array}{ccc}
          \frac{\displaystyle \phi_u}{\displaystyle \sqrt{2}}& 
          \pi^+ & K^+ \\
          \pi^- &
          \frac{\displaystyle \phi_d}{\displaystyle \sqrt{2}} 
          & K^0 \\
          K^- & \bar K^0 &
          \frac{\displaystyle \phi_s}{\displaystyle \sqrt{2}}
          \end{array} \right)
\end{equation}
with the following identifications $\phi_u =\eta_\ns +\pi^0$, 
$\phi_d =\eta_\ns -\pi^0$, $\phi_s =\sqrt{2}\eta_\s$,
$\sigma_u =\epsilon_\ns +a^0_0$, $\sigma_d =\epsilon_\ns -a^0_0$,
and $\sigma_s =\sqrt{2}\epsilon_\s$ for the correctly normalized 
states in the flavour basis (see Eq.(\ref{ortog2}) in Appendix B).
Here the subscripts $\ns$ and $\s$ denote non-strange and strange 
respectively. 

For the set of auxiliary mesonic fields $s,p$ the symmetry 
transformation properties are the same as the ones for $\sigma,\phi$ 
and follow from the chiral transformations of quark fields 
\begin{equation}
\label{fermtr}
   \delta q = i(\alpha+\gamma_5 \beta) q, \qquad 
   \delta {\bar q} = -i  {\bar q}(\alpha-\gamma_5 \beta), 
\end{equation}
where the parameters of the infinitesimal global transformations $\alpha$ 
and $\beta$ are hermitian flavour matrices. One has, for example,
\begin{equation}
\label{inf}
   \delta s =i[\alpha,s]+\{\beta,p\}, \qquad
   \delta p =i[\alpha,p]-\{\beta,s\}.   
\end{equation}
The symmetry breaking piece of the Lagrangian is contained in 
${\cal L}_r$, since    
\begin{equation}
\label{syb}
   \delta  {\cal L}_q = 0, \qquad
   \delta  {\cal L}_r = \delta {\cal L}_\symb \neq 0,
\end{equation}
where
\begin{equation}
\label{sym}
   {\cal L}_{\symb} = -\frac{1}{2}\mbox{tr}(\hat{m}s)
                    +\frac{\kappa}{64}
                    \Big(\det (s+ip)+\det (s-ip)\Big). 
\end{equation}
We see that ${\cal L}_\symb$ is not invariant under a global chiral 
transformation due to explicit symmetry breaking, governed by the 
first term, and due to the 't Hooft interaction, given by the second term
\begin{eqnarray}
\label{symbr}
   \delta {\cal L}_{\symb}\!\!\!\!\!\!\!\! 
   &&=\frac{1}{2}\mbox{tr}\Big(
      i\alpha [\hat{m},s]-\beta \{\hat{m},p\}\Big)
      \nonumber \\
   &&\ \ \ +i\beta_0\frac{\kappa\sqrt 6}{32} 
      \Big(\det (s-ip)-\det (s+ip)\Big). 
\end{eqnarray}
In the following we shall consider the case with diagonal matrix 
$\hat{m}$ where $\hat{m}_u=\hat{m}_d\ne \hat{m}_s$, i.e. the 
chiral symmetry is explicitly broken
down to the vectorial isotopic $SU_I(2)\times U(1)_Y$ symmetry. 
The non-vanishing term proportional to $\kappa$ signals $U_A(1)$ breaking 
leading to the OZI-violating effects related to the Adler -- Bell -- 
Jackiw anomaly of the $SU(3)$ singlet axial current.  

The Fermi fields in Eq.(\ref{lagr2}) enter the action bilinearly and the 
integration over them is exact. The result is given in the next section. 
It is necessary to shift the scalar fields in (\ref{genf3}), 
$\sigma_a(x)\rightarrow\sigma_a(x)+m_a$. It is well known that in
nature the global chiral symmetry $SU_L(3)\times SU_R(3)$ is 
spontaneously broken down to the Eightfold Way symmetry and the shift
takes this into account. In the new vacuum state the vacuum expectation 
values of the shifted fields vanish $\langle 0|\sigma_a(x)|0\rangle =0$. 
The new vacuum is determined by the tadpole mechanism demanding that
all tadpole graphs must sum to zero. The constants $m_a$ denoting 
the constituent quark masses will be fixed by the gap equations. 

In \cite{Reinhardt:1988} the lowest order stationary phase approximation 
(SPA) has been used to estimate the leading contribution from the 't
Hooft determinant in Eq.(\ref{lagr3}) in the functional integrals over 
$s_a$ and $p_a$  
\begin{equation}
\label{intJ}
     {\cal Z}[\sigma +m,\phi ]\equiv 
     {\cal N}\int^{+\infty}_{-\infty}\prod_a{\cal D}s_a{\cal D}p_a
     \exp\left(i\int\! \ud^4 x{\cal L}_r(\sigma +m ,\phi ,s,p)\right),
\end{equation} 
where ${\cal N}$ is chosen such that ${\cal Z}[m,0]=1$.
In the SPA the functional integral is dominated by the
stationary trajectories $r^a_{\st}=(s^a_{\st},\ 
p^a_{\st})$, leading to
\begin{eqnarray}
\label{lagr4}
  && \int\prod_a{\cal D}s_a{\cal D}p_a
     \exp\left(i\int\!\ud^4 x{\cal L}_r(\sigma +m,\phi ,s,p)\right)
     \nonumber \\ 
  &&\simeq\exp\left(i\int\!\ud^4 x{\cal L}_r(r_{\st})\right),
\end{eqnarray}
where $\hbar$ corrections are neglected. The stationary point,
$r^a_{\st}(\sigma ,\phi ;m)$, is a solution of the equations 
${\cal L}'_r(s,p)=0$:
\begin{equation}
\label{saddle}
  \left\{
         \begin{array}{rcr}
        && Gs_a+(\sigma +\Delta )_a
         +\displaystyle\frac{3\kappa}{32}A_{abc}(s_bs_c-p_bp_c)=0. \\
        && Gp_a+\phi_a-\displaystyle\frac{3\kappa}{16}A_{abc}s_bp_c=0,
         \end{array}
  \right.
\end{equation}
where $\Delta_a=m_a - \hat{m}_a$.
This system is well-known from \cite{Reinhardt:1988}. 
Using expressions (\ref{lagr3}) and (\ref{saddle}) we obtain 
\begin{equation}
\label{Lrst}
   {\cal L}_r (r_\st )=\frac{G}{6}\left[
            (s_{\st}^a)^2+(p_{\st}^a)^2\right]
            +\frac{2}{3}\Big( (\sigma +\Delta )_as^a_{\st}
            +\phi_ap^a_{\st}\Big). 
\end{equation}  
One solves Eqs.(\ref{saddle}) exactly, looking for solutions 
$s^a_\st$ and $p^a_\st$ in the form of increasing powers in 
fields $\sigma_a, \phi_a$  
\begin{eqnarray}
\label{rst}
   s^a_\st\!\!\!\!\!\!\!\!
       &&=h_a+h_{ab}^{(1)}\sigma_b
       +h_{abc}^{(1)}\sigma_b\sigma_c
       +h_{abc}^{(2)}\phi_b\phi_c
       +h_{abcd}^{(1)}\sigma_b\sigma_c\sigma_d \nonumber \\
     &&+h_{abcd}^{(2)}\sigma_b\phi_c\phi_d
       +\ldots \nonumber \\
   p^a_\st\!\!\!\!\!\!\!\!
       &&=h_{ab}^{(2)}\phi_b
       +h_{abc}^{(3)}\phi_b\sigma_c
       +h_{abcd}^{(3)}\sigma_b\sigma_c\phi_d
       +h_{abcd}^{(4)}\phi_b\phi_c\phi_d
       +\ldots 
\end{eqnarray}
with coefficients depending on $m_a$ and coupling constants.
Putting these expansions in Eqs.(\ref{saddle}) one obtains a series 
of selfconsistent equations to determine coefficients $h_a$, $h^{(1)}_{ab}$, 
$h^{(2)}_{ab}$ and so on. The first three of them are
\begin{eqnarray}
\label{ha}
   &&Gh_a+\Delta_a+\frac{3\kappa}{32}A_{abc}h_bh_c=0, \nonumber \\  
   &&\left(G\delta_{ac}+\frac{3\kappa}{16}A_{acb}h_b
     \right)h^{(1)}_{ce}=-\delta_{ae}\ , \\
   &&\left(G\delta_{ac}-\frac{3\kappa}{16}A_{acb}h_b
     \right)h^{(2)}_{ce}=-\delta_{ae}\ . \nonumber
\end{eqnarray}
All the other equations can be written in terms of the already known 
coefficients, for instance, we have \cite{Osipov7:2002}
\begin{eqnarray}
\label{coeffi}
   &&h^{(1)}_{abc}
     =\frac{3\kappa}{32}h^{(1)}_{a\bar a}h^{(1)}_{b\bar b} 
     h^{(1)}_{c\bar c}A_{\bar a\bar b\bar c}\ , \quad \ \  
     h^{(2)}_{abc}=-\frac{3\kappa}{32}h^{(1)}_{a\bar a}h^{(2)}_{b\bar b} 
     h^{(2)}_{c\bar c}A_{\bar a\bar b\bar c}\ , \nonumber \\ 
   &&h^{(3)}_{abc}
     =-\frac{3\kappa}{16}h^{(2)}_{a\bar a}h^{(2)}_{b\bar b} 
     h^{(1)}_{c\bar c}A_{\bar a\bar b\bar c}\ , \quad 
     h_{abcd}^{(1)}=\frac{3\kappa}{16}h^{(1)}_{a\bar a}
     h^{(1)}_{b\bar b}h^{(1)}_{\bar ccd}
     A_{\bar a\bar b\bar c}\ , \nonumber \\ 
   &&h_{abcd}^{(2)}
     =\frac{3\kappa}{16}h^{(1)}_{a\bar a}
     \left(h^{(1)}_{b\bar b}h^{(2)}_{\bar ccd}
     -h^{(2)}_{c\bar b}h^{(3)}_{\bar cdb}\right)
     A_{\bar a\bar b\bar c}\ , \quad\ldots 
\end{eqnarray}
One can see from these equations that the terms quadratic and higher 
order in mesonic fields in Eqs.(\ref{rst}) are generated by the 't 
Hooft interaction and will disappear if $\kappa =0$. Let us also give 
the relations following from (\ref{ha}) which have been used to obtain 
(\ref{coeffi}) 
\begin{equation}
  h_b=(Gh_a+2\Delta_a)h^{(1)}_{ab}
     =-(3Gh_a+2\Delta_a)h^{(2)}_{ab}.
\end{equation}

As a result the effective Lagrangian (\ref{Lrst}) can be expanded in powers 
of meson fields. Such an expansion, up to and including the terms
which are cubic in $\sigma_a, \phi_a$, looks like
\begin{eqnarray}
\label{lam}
   {\cal L}_r(r_\st )\!\!\!\!\!\!\!\!
   &&=h_a\sigma_a+\frac{1}{2}h_{ab}^{(1)}\sigma_a\sigma_b  
      +\frac{1}{2}h_{ab}^{(2)}\phi_a\phi_b \nonumber \\
   &&\ \ \ +\frac{1}{3}\sigma_a\left[h^{(1)}_{abc}\sigma_b\sigma_c
   +\left(h^{(2)}_{abc}+h^{(3)}_{bca}\right)\phi_b\phi_c\right]
   +{\cal O}(\mbox{field}^4).
\end{eqnarray} 

The coefficients $h_a$ are determined by couplings $G, \kappa$
and the mean field $\Delta_a$. This field has in general only three 
non-zero components with indices $a=0,3,8$, according to the symmetry 
breaking pattern. The same is true for $h_a$ because of the first 
equation in (\ref{ha}). It means that there is a system of only three 
equations to determine $h=h_a\lambda_a=\mbox{diag}(h_u,h_d,h_s)$, 
\begin{equation}
\label{order-h0}
   \Delta_i+Gh_i+\frac{\kappa}{32}\sum_{j,k}t_{ijk}h_jh_k=0. 
\end{equation}
Here the totally symmetric coefficients $t_{ijk}$ are zero except for 
the case with different values of indices $i\neq j\neq k$ when 
$t_{uds}=1$. The latin indices $i,j,k$ mark the flavour states 
$i=u,d,s$ which are linear combinations of states with indices $0,3$ 
and $8$. In Appendix A we collect the matrices which project one set 
to the other and write out exact solutions for Eq.(\ref{ha}). Let us
note that Eqs.(\ref{order-h0}) must be solved selfconsistently with 
the gap equations (see Eq.(\ref{gap}) below) to yield the constituent 
quark masses in leading SPA order. 


\section*{\normalsize 3. Heat kernel expansion}

Equation (\ref{lam}) contains the piece of the bosonized effective 
Lagrangian, which has no kinetic terms and is obtained in the weak 
field limit. Now we turn to the evaluation of the fermionic functional 
integral in Eq.(\ref{genf3}), which after the shift 
$\sigma_a(x)\rightarrow\sigma_a(x)+m_a$, reads
\begin{equation}
\label{Zferm}
   Z[Y]=\int {\cal D}q{\cal D}\bar{q}\exp\Big(
        i\int\!\ud^4x\bar{q}[i\gamma^\mu\partial_\mu -(m+\sigma 
        +i\gamma_5\phi )]q\Big),
\end{equation}
where $Y$ collects the background field dependence as indicated below.
This fermion determinant accounts for the remaining part of the effective
Lagrangian and leads, in general, to nonlocal mesonic vertices with
unphysical cuts (the quark deconfinement problem). We have resorted here 
to the Schwinger -- DeWitt representation for the real part of the 
corresponding effective action, $W[Y]$, to obtain in the end the  
asymptotics for $W[Y]$ in terms of local polynomials of 
background fields and their derivatives given by the heat kernel 
coefficients at coinciding arguments, 
\begin{eqnarray}
   Z[Y]\!\!\!\!\!\!\!\!
       &&=\exp(W[Y]), \nonumber \\ 
   W[Y]\!\!\!\!\!\!\!\!
       &&=\ln|\det D|=-\frac{1}{2}\int_{0}^{\infty}
          \frac{\ud t}{t}\rho (t\Lambda^2)\mbox{Tr}\exp 
          (-t{D^{\dagger}_\e D}_\e ),
\end{eqnarray}
where $\mbox{Tr}$ designates functional trace, the operator $D_\e$ stands 
for the euclidean Dirac operator in presence of the background fields 
$\sigma,\phi$ and
\begin{equation}
   {D^{\dagger}_\e D}_\e =m^2-\partial^2+Y,
\end{equation}
with the definition
\begin{equation}
   Y=i\gamma_{\mu}(\partial_{\mu} \sigma+i\gamma_5\partial_{\mu}\phi)
    +\sigma^2+\{m,\sigma\}+\phi^2+i\gamma_5[\sigma+m,\phi ].
\end{equation}

For the regulator $\rho (t\Lambda^2)$, needed to keep the integral 
convergent at $t=0$, we use two Pauli-Villars subtractions\footnote{A
regularization function $\rho$ must be introduced to define the 
coincidence limit for the Schwinger representation. The
regularization of the quark determinant in general should be done in 
accordance with certain requirements (see, for example, the review of 
R.D. Ball in \cite{Vassilevich:2003}). Some of them are discussed also
in \cite{Diakonov:1989}.} 
\begin{equation}
   \rho (t\Lambda^2)=1-(1+t\Lambda^2)\mbox{exp}(-t\Lambda^2),
\end{equation}
where the cut-off $\Lambda$ is a free dimensionfull parameter. 
The regularization function $\rho (t\Lambda^2)$, being written 
in terms of a dimensionless variable $\tau =t\Lambda^2$, fulfills 
the necessary conditions: $\rho (\tau )\sim \tau^2/2$ at 
$\tau\rightarrow 0$ and $\rho (\tau )\rightarrow 1$ at $\tau
\rightarrow\infty$. It is important to know to what extent the 
specific form of this function affects our results.   
It is obvious that the type of used regulator does not affect 
the chiral invariance of the heat kernel expansion, since the 
generalized heat kernel coefficients $b_i$ \cite{Osipov2:2001}, 
which carry the whole symmetry properties of the heat kernel 
expansion, do not depend on it
\begin{equation}
\label{ZY}
   W[Y]=-\int\frac{\ud^4 x_\e}{32\pi^2}\sum_{i=0}^{\infty}
        I_{i-1}\mbox{tr}(b_i).
\end{equation}
Here the expressions for the first four $b_i$ in the case of 
$SU(2)_I\times U(1)_Y$ flavour symmetry $m_u=m_d\ne m_s$ are 
\begin{eqnarray}
\label{SW}
   b_0\!\!\!\!\!\!\!\!
      &&=1,  \nonumber\\
   b_1\!\!\!\!\!\!\!\!
      &&=-Y, \nonumber\\
   b_2\!\!\!\!\!\!\!\!
      &&=\frac{Y^2}{2}+\frac{\Delta_{us}}{\sqrt{3}}\lambda_8 Y,
         \nonumber\\
   b_3\!\!\!\!\!\!\!\!
      &&=-\frac{Y^3}{3!}+\frac{\Delta_{us}^2}{6\sqrt{3}}
      \lambda_8 Y-\frac{\Delta_{us}}{2\sqrt{3}}\lambda_8 Y^2
      -\frac{1}{12}(\partial Y)^2,
\end{eqnarray}
where we used the definition $\Delta_{ij}\equiv m_i^2-m_j^2$.
In (\ref{ZY}) the trace is to be taken over colour, flavour and Dirac 
4-spinors indices and the regulator-dependent integrals $I_i$ are 
the weighted sums \cite{Osipov2:2001}
\begin{equation}
   I_i=\frac{1}{3}\left(2J_i(m_u^2)+J_i(m_s^2)\right)
\end{equation}
with 
\begin{equation}
\label{Ji}
   J_i(m_j^2)=\int_{0}^{\infty}\frac{\ud t}{t^{2-i}}\rho (t\Lambda^2)
              \exp (-tm_j^2).
\end{equation}
For the chosen form of the cut-off function we obtain, for instance,
\begin{eqnarray}
\label{J0}
   J_0(m^2)\!\!\!\!\!\!\!\!
     &&=\Lambda^2-m^2\ln\left(1+\frac{\Lambda^2}{m^2}\right), \\
   J_1(m^2)\!\!\!\!\!\!\!\!
     &&=\ln\left(1+\frac{\Lambda^2}{m^2}\right)
             -\frac{\Lambda^2}{\Lambda^2+m^2}.
\end{eqnarray}
Both of them are divergent in the limiting case $\Lambda\rightarrow\infty$.

Thus, the effective Lagrangian depends on the integrals $I_i$.
The more terms of the heat kernel series are taken into account, the 
more the final result depends on the form of the cut-off function 
$\rho (\tau )$ and, therefore, the more careful one should be choosing 
a regulator. In the following we restrict our study to the two nontrivial 
terms, $b_1$ and $b_2$, in the asymptotic expansion of $W[Y]$. In this 
case only two integrals, $I_0$ and $I_1$, are involved. If we 
introduced in $\rho (\tau)$ two independent parameters, instead of one, 
$\Lambda$, the outcome would not depend at all on the form of the
regulator, because one can always fix these parameters by fixing 
independently couplings $I_0$ and $I_1$ from experimental data. 
Actually, we slightly simplified our calculations working with only
one parameter $\Lambda$, paying for that the price of having some 
dependence on the regularization procedure, which is finally inherited 
by the constituent quark masses.   

The heat kernel series (\ref{ZY}) defines the asymptotics of the 
effective action for a physical system with the mass matrix $m$
being large compared to the rest of the background fields and
their derivatives. It corresponds exactly to the considered case
of low-energy QCD, where the small meson exitations of the quark 
sea take place in the ''superconducting'' phase with heavy 
constituent quarks. It is interesting to stress that in comparison 
with the standard Seeley -- DeWitt coefficients, which transform 
covariantly with respect to the action of the chiral group, our 
coefficients $b_i$ possess more specific transformation properties. 
Indeed, in the broken vacuum state an arbitrary
infinitesimal variation $\delta\mbox{tr}(b_i)$, induced by global
transformations of the background fields 
\begin{eqnarray}
\label{inf2}
   \delta \sigma\!\!\!\!\!\!\!\!
   &&=i[\alpha,\sigma +m]+\{\beta,\phi\}, \nonumber \\
   \delta \phi\!\!\!\!\!\!\!\!
   &&=i[\alpha,\phi ]-\{\beta,\sigma +m\},   
\end{eqnarray}
depends on the variation $\delta Y$ which is equal to
\begin{equation} 
\label{deltaY}
   \delta Y\!=i[\alpha +\gamma_5\beta,Y+m^2].
\end{equation}
One can see that already the first coefficient $b_1$ transforms
non-covariantly, because $m^2$ does not commute with $\alpha +
\gamma_5\beta$ in (\ref{deltaY}). Nevertheless, one can prove that  
$\delta\mbox{tr}(b_i)=0$ for all generalized coefficients $b_i$
\cite{Osipov2:2001}. 

In the present calculations we truncate the heat kernel series at $b_2$. 
In this approximation the effective Lagrangian ${\cal L}$ is given by 
the sum of only two local terms ${\cal L}={\cal L}(b_1)+{\cal L}(b_2)+
\ldots,$ where 
\begin{eqnarray}
\label{tL}
   {\cal L}(b_1)\!\!\!\!\!\!\!\!
   &&={\cal L}_\tad (b_1)+{\cal L}_\mass (b_1), \nonumber \\
   {\cal L}(b_2)\!\!\!\!\!\!\!\!
   &&={\cal L}_\tad (b_2)+{\cal L}_\kin (b_2)
     +{\cal L}_\mass (b_2)+{\cal L}_\inter (b_2).
\end{eqnarray}
Here we distinguish the tadpole terms, ${\cal L}_\tad$, from mass terms,
${\cal L}_\mass $, kinetic terms, ${\cal L}_\kin $, and interaction terms,
${\cal L}_\inter $. We have, for instance,
\begin{eqnarray}  
\label{tadpole}
   {\cal L}_\tad (b_1)\!\!\!\!\!\!\!\!
   &&=\frac{N_cI_0}{4\pi^2}[m_u(\sigma_u +\sigma_d)+m_s\sigma_s],
      \nonumber \\
   {\cal L}_\tad (b_2)\!\!\!\!\!\!\!\!
   &&=-\frac{N_cI_1}{12\pi^2}\Delta_{us}[m_u(\sigma_u +\sigma_d)
      -2m_s\sigma_s].
\end{eqnarray}
Joined together with the tadpole contribution from Lagrangian (\ref{lam}), 
they lead to the gap equations
\begin{equation}
\label{gap}
  \left\{
       \begin{array}{rcr}
  && h_u+\displaystyle\frac{N_c}{6\pi^2} m_u
         \left(3I_0-\Delta_{us} I_1 \right)=0. \\
  && \\
  && h_s+\displaystyle\frac{N_c}{6\pi^2} m_s
         \left(3I_0+2\Delta_{us} I_1 \right)=0.
        \end{array}
  \right.
\end{equation}

The mass-part of the heat kernel effective Lagrangian contains two 
contributions and is given by
\begin{eqnarray}
\label{mass}
     {\cal L}_\mass^{(b_1+b_2)}
     \!\!\!\!\!\!\!\!\!
     &&=\frac{N_cI_0}{4\pi^2}\left( \sigma_a^2+\phi_a^2\right)  
        -\frac{N_cI_1}{12\pi^2}
        \left\{
        \Delta_{us}[2\sqrt{2}(3\sigma_0\sigma_8+\phi_0\phi_8)
        -\phi_8^2+\phi_i^2]
        \right.\nonumber \\
     &&\ \ \ +2(2m_u^2+m_s^2)\sigma_0^2+(m_u^2+5m_s^2)\sigma_8^2
       +(7m_u^2-m_s^2)\sigma_i^2 \nonumber \\
     &&\ \ \ \left. +(m_u+m_s)(m_u+2m_s)\sigma_f^2
       +(m_s-m_u)(2m_s-m_u)\phi_f^2\right\},
\end{eqnarray}
where we assume that the indices $i$ and $f$ range over the subsets
$i=1,2,3$ and $f=4,5,6,7$ of the set $a=0,1,\ldots,8.$ Thus we have
\begin{eqnarray}
\label{not}
   && \phi_i^2=2\pi^+\pi^-+(\pi^0)^2,\qquad 
      \phi_f^2 = 2(K^+K^-+\bar{K}^0K^0), \nonumber \\
   && \sigma_i^2=2a_0^+a_0^-+(a_0^0)^2,\qquad 
   \sigma_f^2 = 2(K^{*+}_0 K^{*-}_0 + \bar{K}^{*0}_0 
   K^{*0}_0). 
\end{eqnarray}

The kinetic term, ${\cal L}_\kin (b_2)$, after continuation to Minkowski
space, has a non-standard factor 
\begin{equation}
\label{kinn}
   {\cal L}_\kin (b_2)=\frac{N_cI_1}{16\pi^2}\mbox{tr}\left[
                       (\partial_\mu \sigma )^2+
                       (\partial_\mu \phi )^2\right]. 
\end{equation}
It should be rescaled by the redefinition of mesonic fields
\begin{equation}
\label{renorm}  
   \sigma_a = g \sigma_a^\ren , \quad 
   \phi_a = g \phi_a^\ren , \quad
   g^2=\frac{4\pi^2}{N_cI_1}\ .
\end{equation}
where the index $\ren$ stands for the new renormalized fields. 

By virtue of the PCAC hypothesis the coupling $g$ is related to 
the weak decay constants of the pion, $f_\pi$, or the kaon, $f_K$, 
\begin{equation}
\label{fpi}
   f_\pi =\frac{m_u}{g}\ , \qquad f_K=\frac{m_s+m_u}{2g}\ .
\end{equation}
To see this let us recall Eq.(\ref{1}), where the quarks bilinears
$\bar{q}\lambda_aq$ and $\bar{q}i\gamma_5\lambda_aq$ have been
replaced by the auxiliary fields $s_a$ and $p_a$. The SPA
aproximation used to estimate the path integral over these variables
in (\ref{lagr4}) restricts them to the stationary trajectories 
$s^a_\st , p^a_\st$, given by Eq.(\ref{rst}). Thus, we have 
\begin{equation}  
\label{qb}
   i\bar{q}\gamma_5\lambda_aq=p_a^\st , \qquad
   \bar{q}\lambda_aq=s_a^\st .  
\end{equation}
The quark operators are finally represented by expansions in 
increasing powers of bosonic fields $\sigma_a$ and $\phi_a$. This 
is a convenient form to establish a connection to some current algebra 
results, such as the PCAC hypothesis or the Gell-Mann -- Oakes -- Renner 
(GOR) relation \cite{Gell-Mann:1968}. 

For instance, one easily finds from (\ref{qb}), 
\begin{equation} 
   \big <\pi^-|\bar{d}\gamma_5u|0\big >=
        \frac{ig\big <\pi^-|\phi_{\pi^+}^\ren |0\big >}{\sqrt{2}
        G(1+\omega_s)}
        =\frac{im_\pi^2}{\sqrt{2}\hat{m}_u}
        \left(\frac{m_u}{g}\right)\big <\pi^-|\phi_{\pi^+}^\ren 
        | 0\big >,
\end{equation}
where result (\ref{mpion}) has been used to obtain the last equality. 
In exactly the same way one derives with the help of Eq.(\ref{K-mass}) 
\begin{equation} 
   \big <K^-|\bar{s}\gamma_5u|0\big >=
   \frac{ig\big <K^-|\phi_{K^+}^\ren |0\big >}{\sqrt{2}G(1+\omega_u)}
        =\frac{i\sqrt{2}m_K^2}{(\hat{m}_u+\hat{m}_s)}
        \left(\frac{m_u+m_s}{2g}\right)\big <K^-|\phi_{K^+}^\ren 
        |0\big >.
\end{equation}
Let us assume that (\ref{fpi}) holds, then these equations coincide
with the well known PCAC relations. 

One can use the second equation in (\ref{qb}) to estimate the quark
condensates in the vacuum. As far as the isotopic invariance is 
implemented here we have
\begin{equation}
   \big <0|\bar{u}u|0\big >=\big <0|\bar{d}d|0\big >=\frac{h_u}{2}\ ,
   \qquad  
   \big <0|\bar{s}s|0\big >=\frac{h_s}{2}\ . 
\end{equation}
Combining these equations with Eqs.(\ref{mpion}), (\ref{K-mass}) and
(\ref{fpi}) one finds the GOR relations (up to the last terms in the
round brackets, which are proportional to the current quark masses and
give some model corrections to the leading order result)
\begin{eqnarray}
   &&m_\pi^2f_\pi^2=-2\hat{m}_u\big <0|\bar{u}u|0\big >
                    \left(1+\frac{\hat{m}_u}{\Delta_u}\right). \\
   &&m_K^2f_K^2=-\frac{1}{2}(\hat{m}_u+\hat{m}_s)
                \big <0|\bar{u}u+\bar{s}s|0\big >
                \left(1+\frac{\hat{m}_u+\hat{m}_s}{
                \Delta_u+\Delta_s}\right).
\end{eqnarray}


\section*{\normalsize 4. Mass spectrum}

We proceed now to extract the mass terms for the low-lying pseudoscalar 
and scalar nonets. We discuss first the pseudoscalar spectrum. The quadratic 
terms in the fields from Eq.(\ref{lam}) and Eq.(\ref{mass}) combine to
yield for instance
\begin{eqnarray}
   {\cal L}_\mass (\pi )\!\!\!\!\!\!\!\!\!
   &&=\phi_i^2\left[\frac{N_c}{12\pi^2}\left( 3I_0-\Delta_{us}I_1 \right)
      -\frac{1}{2G(1+\omega_s)}\right] 
      \nonumber \\
   &&=-\frac{\hat{m}_u\phi_i^2}{2Gm_u(1+\omega_s)}\ .
\end{eqnarray} 
To get this result we used the gap equation (\ref{gap}) and the
stationary phase conditions (\ref{order-h0}). Let us also remind that 
some of our notations and results are explained in Appendix A. Finally 
the pion mass is obtained by introducing physical fields (\ref{renorm})
\begin{equation}
\label{mpion}
   m_\pi^2=\frac{g^2\hat{m}_u}{Gm_u(1+\omega_s)}\ .
\end{equation}   

In exactly the same way one can obtain the masses of the other members 
of the pseudoscalar nonet 
\begin{eqnarray}
\label{K-mass}
   m_K^2\!\!\!\!\!\!\!\!\!
   &&=\frac{g^2(\hat{m}_u+\hat{m}_s)}{G(m_u+m_s)(1+\omega_u)}\ , \\ 
   m^2_\eta\!\!\!\!\!\!\!\!\!
   &&=\frac{g^2}{2}\left(
      A+B-\sqrt{(A-B)^2+4D^2}\right), \\
   m^2_{\eta '}\!\!\!\!\!\!\!\!\!
   &&=\frac{g^2}{2}\left(
            A+B+\sqrt{(A-B)^2+4D^2}\right).
\end{eqnarray}
We also have     
\begin{eqnarray}
   A+B\!\!\!\!\!\!\!\!\!
   &&=\frac{h_u}{m_u}+\frac{h_s}{m_s}+
      \frac{2-\omega_s}{G\mu_-}\ ,
      \nonumber \\
   A-B\!\!\!\!\!\!\!\!\!
   &&=\frac{1}{3}\left(\frac{h_u}{m_u}-\frac{h_s}{m_s}+
      \frac{8\omega_u+\omega_s}{G\mu_-}\right),
      \nonumber \\
   D\!\!\!\!\!\!\!\!\!
   &&=\frac{\sqrt{2}}{3}\left(\frac{h_u}{m_u}-\frac{h_s}{m_s}+
      \frac{\omega_s-\omega_u}{G\mu_-}\right),
\end{eqnarray}
where $\mu_\pm =(1\pm\omega_s-2\omega_u^2)$. The argument of the square 
root is
\begin{equation}
   (A-B)^2+4D^2=\left( \frac{h_u}{m_u}-\frac{h_s}{m_s}+
                \frac{\omega_s}{G\mu_-}\right)^2
                +8\left(\frac{\omega_u}{G\mu_-}\right)^2.
\end{equation}
 
It is known that for $m_u=m_d\ne m_s$ there is mixing in the $0,8$ 
channels. This part of the Lagrangian has been diagonalized by introducing
physical fields $\eta$ and $\eta'$ via an orthogonal transformation,
as it is discussed in Appendix B, with the mixing angle $\theta_\p$ 
(in the singlet -- octet basis) defined from the diagonalization 
requirement.

In the limit of vanishing 't Hooft interaction, $\kappa=0$, the 
mixing angle $\theta_\p$ is equal to the ideal one with $\tan 
(2\theta_\id )=2\sqrt{2}$ and one can conclude that 
$\eta\sim\eta_\ns,\ \eta'\sim -\eta_\s$. We find in this case
\begin{equation}
\label{mpi0}
   m_\pi^2=m^2_{\eta_\ns }=
           \frac{g^2\hat{m}_u}{Gm_u}\ , \qquad
   m_K^2=\frac{g^2(\hat{m}_u+\hat{m}_s)}{G(m_u+m_s)}\ , \qquad  
   m^2_{\eta_\s }=\frac{g^2\hat{m}_s}{Gm_s}\ .
\end{equation}
Using the gap equations one obtains the relations
\begin{equation}
   \frac{ m_{K}^2-m_{\pi}^2}{2m_u(m_s-m_u)}=\frac{m_s}{m_u}
     \ , \qquad
   \frac{m_{\eta_\s}^2-m_{K}^2}{2m_u(m_s-m_u)}=1,
\end{equation}
which show the mass splittings within the nonet.

In the $SU(3)$ limit $m_u=m_d=m_s$ for non-vanishing $\kappa$ 
there is no $\phi_0-\phi_8$ mixing, since $D=0$. One obtains 
immediately the masses 
\begin{equation}
\label{mpik}
   m_{\pi}^2=m^2_K=m^2_{88}= 
   \frac{g^2\hat{m}_u}{G m_u (1+\omega )}\ , 
\end{equation}
with the singlet-octet mass splitting  
\begin{equation}
\label{so}
   m_{00}^2-m_{88}^2=\frac{3g^2\omega}{G(1+\omega )(1-2\omega )}\ ,
\end{equation}
where 
\begin{equation}
\label{omegasu3}
   \omega =\frac{\kappa h}{16G}=\frac{1}{2}\left(
           \sqrt{1-\frac{\kappa\Delta_u}{4G^2}}-1\right)
\end{equation}
is a solution of the stationary phase equation (\ref{order-h0}) for the 
$SU(3)$ case. In the chiral limit, $\hat{m}=0$, the singlet mass $m_{00}$  
takes a non-vanishing value. The would-be $U(1)$ Goldstone boson receives 
a mass as a result of the 't Hooft interaction. 

We turn now to the scalar sector. The masses of the scalar mesons are as 
follows. For the mesons usually refered to as $a_0$ 
$(I^G(J^{PC})=1^-(0^{++}))$ we obtain 
\begin{equation}
\label{si}
  m_{a_0}^2=g^2\left(\frac{h_u}{m_u}+\frac{1}{G(1-\omega_s)}\right)
               +4m_u^2
              =m_\pi^2+4m_u^2+\frac{2g^2\omega_s}{G(1-\omega_s^2)}\ .
\end{equation}
and for the strange $K_0^*$ $(I(J^P)=\frac{1}{2}(0^+))$ we have
\begin{eqnarray}
\label{de}
   m_{K^{*}_0}^2\!\!\!\!\!\!\!\!\! 
     &&=g^2\left(\frac{1}{G(1-\omega_u)}
        +\frac{h_u+h_s}{m_u+m_s} \right)
        +4m_sm_u \nonumber \\
     &&=m_K^2+4m_sm_u+\frac{2g^2\omega_u}{G(1-\omega_u^2)}\ .
\end{eqnarray}

In the $0,8$ channels one must diagonalize the states. 
Diagonalization proceeds as in the pseudoscalar case and 
the resulting scalar states are denoted by $\epsilon$ and 
$\epsilon'$ respectively indicating a set of $f_0$ 
$(I^G(J^{PC})=0^+(0^{++}))$ mesons.
The mixing angle $\theta_\s$ is defined in the $(0,8)$ basis.
As a result we obtain for the corresponding masses
\begin{eqnarray}
   m^2_\epsilon\!\!\!\!\!\!\!\!\! 
   &&=\frac{g^2}{2}\left(
      {\cal A}+{\cal B}-\sqrt{({\cal A}-{\cal B})^2+4{\cal D}^2}\right),
      \nonumber \\
   m^2_{\epsilon '}\!\!\!\!\!\!\!\!\! 
   &&=\frac{g^2}{2}\left(
      {\cal A}+{\cal B}+\sqrt{({\cal A}-{\cal B})^2+4{\cal D}^2}\right),
\end{eqnarray}
where  
\begin{eqnarray}
   {\cal A+B}\!\!\!\!\!\!\!\!\!
   &&=\frac{h_u}{m_u}+\frac{h_s}{m_s}
      +\frac{N_cI_1}{\pi^2}\left(m_s^2+m_u^2\right)
      +\frac{2+\omega_s}{G\mu_+}\ , 
      \nonumber \\
   {\cal A-B}\!\!\!\!\!\!\!\!\!
   &&=\frac{h_u}{m_u}-\frac{h_s}{m_s}
      -\frac{8\omega_u+\omega_s}{3G\mu_+}\ ,
      \nonumber \\
   {\cal D}\!\!\!\!\!\!\!\!\!
   &&=\sqrt{2}\left(\frac{h_u}{m_u}-\frac{h_s}{m_s}-
     \frac{\omega_s-\omega_u}{3G\mu_+}\right),
\end{eqnarray}
and 
\begin{equation}
  ({\cal A-B})^2+4{\cal D}^2=\left[3\left(\frac{h_u}{m_u}
                            -\frac{h_s}{m_s}\right)-
  \frac{\omega_s}{G\mu_+}\right]^2+
  8\left(\frac{\omega_u}{G\mu_+}\right)^2.
\end{equation}

Supposing for a moment that $\kappa=0$, we find the mixing angle 
$\theta_\s$ to be equal $\theta_\id$, the $\epsilon$-meson is  
a pure non-strange state, $\epsilon_\ns$, and the $\epsilon'$ is purely
strange, $-\epsilon_\s$.  The scalar masses become
\begin{eqnarray}
\label{skappa0}
   &&m_{a_0}^2=m_{\epsilon_\ns}^2=m_\pi^2+4m_u^2, \nonumber \\
   &&m_{K^{*}_0}^2=m_K^2+4m_um_s, \nonumber\\
   &&m_{\epsilon_\s}^2=m_{\eta_\s}^2+4m_s^2,
\end{eqnarray}
giving the following mass splittings within the nonet
\begin{eqnarray}
   &&m_{K^{*}_0}^2-m_{a_0}^2=2(m_s-m_u)(m_s+2m_u), \nonumber \\
   &&m_{\epsilon_\s}^2-m_{K^{*}_0}^2=2(m_s-m_u)(2m_s+m_u).
\end{eqnarray}
The latter is three times bigger than in the pseudoscalar case
\begin{equation} 
\label{pssos} 
   m^2_{\epsilon_\s}-m^2_{\epsilon_\ns}=3(m^2_{\eta_\s}-m^2_{\eta_\ns} )
   =6(m_s^2-m_u^2).
\end{equation}

Let us consider now the $SU(3)$ limit $m_u=m_d=m_s$ for $\kappa\ne 0$. 
One has 
\begin{equation}
\label{soctet}
   m^2_{a_0}=m^2_{K^{*}_0}=M^2_{88} 
   =m_\pi^2+4m_u^2
   +\frac{2g^2\omega}{G(1-\omega^2)}\ . 
\end{equation}
There is no mixing here, since ${\cal D}=0$, and the singlet state is 
splitted due to the 't Hooft interaction  
\begin{equation}
\label{sos}
   M^2_{00}-M^2_{88}
   =-\frac{3g^2\omega}{G(1-\omega)(1+2\omega )}\ .
\end{equation}
Comparing the $SU(3)$ limit of singlet-octet mass splittings in the 
pseudoscalar, Eq.(\ref{so}), and scalar, Eq.(\ref{sos}), channels, one 
observes that these expressions have opposite signs for the physically 
reasonable sets of parameters ($0<\omega<1/2$), where $\mu_-$ and
$\mu_+$ are positive. The 't Hooft interaction pulls the singlet
pseudoscalar state up and the singlet scalar state down with respect
to the corresponding octet ones.
 
To summarize, the pseudoscalar and scalar masses are obtained by
means of a specific asymptotic expansion\footnote{
A summation over all constant meson fields in this series leads to a
derivative (long wavelength) expansion.} of the heat kernel in the 
framework of a simple model for low energy QCD. It can be
improved in different ways. We have already mentioned some of them 
in Section 2. Here we also would like to point out that in truncating 
the heat kernel series at second order we are neglecting finite 
size momentum dependent contributions to the one-loop fermion 
determinant that become more important for the heavier particles, so
that the pole position for extraction of the masses can be modified in
a sizeable way. However it is well known that the lack of confinement
in the NJL model introduces serious difficulties with the crossing of
non-physical thresholds associated with the production of free 
quark -- antiquark pairs, which one may encounter by formally 
continuing the full Euclidean action to Minkowski space. 
   These are the main reasons why we decided in this simplified version
of the model to truncate the series, taking into account only the 
divergent contributions. On one hand, in doing so, we admittedly
deviate from the original NJL Lagrangian, however in a way which 
relies heavily on its symmetries and asymptotic dynamics, which are 
fully taken into account. On the other hand, this approach gives us, 
in principle, a chance to correct systematically the coefficients 
$I_i$ of the heat kernel series by introducing new parameters 
in the regularization function $\rho(t,\Lambda_1,\Lambda_2, \ldots )$ 
and fixing them in accordance with phenomenological requirements.    
This procedure, hopefully, can be developed similary to QCD sum 
rules, like it has been done in \cite{Shifman:1979} and discussed, 
in particular, in relation with NJL-type models in \cite{Broniowski:1999}.


\section*{\normalsize 5. Numerical results and discussion}

The parameters of the model, $\hat{m}_u$,  $\hat{m}_s$, $G$, $\kappa$ and 
$\Lambda$ are shown in Table 1.   
\vspace{0.5cm}

\noindent {\small Table 1 \\
The main parameters of the model given in the following units: 
$[m]=\mbox{MeV}$, $[G]=\mbox{GeV}^{-2}$, $[\kappa ]=\mbox{GeV}^{-5}$, 
$[\Lambda ]=\mbox{GeV}$.} \\
\begin{center}
\begin{tabular}{|l|cc|cc|c|c|c|c|c}
      \hline
      & $\hat{m}_u$ 
      & $(m_u) $ 
      & $\hat{m}_s$
      & $(m_s) $
      & $G$  
      & $-\kappa $ 
      & $\Lambda $  \\ 
\hline
\hline
a &4.9  &(302)   &167   &(519)   &9.3   &0*   &0.95 \\ \hline
b &2.8  &(211)   &85    &(356)   &2.8   &157  &1.4  \\ \hline
c &2.7  &(214)   &92    &(397)   &3.1   &88   &1.4  \\ \hline
d &1.2  &(171)   &41    &(310)   &1.1   &11   &2.3  \\ \hline
e &0.7  &(155)   &24    &(296)   &0.6   &1.6  &3.2  \\ \hline
f &3.2  &(227)   &105   &(405)   &3.7   &173  &1.3  \\ \hline
g &4.9  &(296)   &161   &(493)   &7.6   &664  &0.95 \\ \hline
h &2.2  &(199)   &75    &(375)   &2.3   &45   &1.6  \\ \hline
i &3.6  &(242)   &122   &(437)   &4.6   &205  &1.2  \\ \hline
j &3.6  &(235)   &109   &(382)   &3.7   &422  &1.2  \\ \hline
k &4.7  &(286)   &155   &(485)   &7.2   &477  &0.98 \\ \hline
l &1.5  &(179)   &50    &(317)   &1.5   &23.4 &2.0  \\ \hline
\end{tabular}
\end{center}   
\vspace{1cm}

\noindent {\small Table 2.
\\
The pseudoscalar nonet parameters in units of MeV (except for the angle
$\theta_\p$, which is given in degrees).}
\\ 
\begin{center}
\begin{tabular}{|l|c|c|c|c|c|c|c|}
\hline
   &$m_\pi$
   &$m_K  $
   &$f_\pi$
   &$f_K  $
   &$m_\eta $
   &$m_{\eta'}$
   &$\theta_\p\ $ \\ 
\hline
\hline
a &138*   &494*  &92*  &125* &138  &612  &35   \\ \hline
b &138*   &494*  &92*  &124  &547* &1504 &2    \\ \hline
c &138*   &494*  &92*  &131  &526  &958* &-4   \\ \hline
d &138*   &494*  &92*  &129  &547* &1078 &2    \\ \hline
e &138*   &495*  &92*  &134  &545* &958* &2    \\ \hline
f &137*   &496*  &92*  &128  &532  &1109 &-2   \\ \hline
g &137*   &496*  &92*  &122* &507  &1089 &-7   \\ \hline
h &138*   &495*  &92*  &133  &535  &958* &-3*  \\ \hline
i &138*   &495*  &92*  &129  &516  &958* &-7*  \\ \hline
j &138*   &494*  &92*  &121* &547* &2187 &2    \\ \hline
k &138*   &494*  &92*  &124* &497  &958* &-10  \\ \hline
l &138*   &494*  &92*  &127* &547* &1156 &2    \\ \hline
\end{tabular}
\end{center}
\vspace{0.5cm}

In Table 2 is the pseudoscalar 
spectrum, together with the weak decay constants $f_\pi$, $f_K$  and 
mixing angle $\theta_\p$;  the masses and mixing angle $\theta_\s$ 
of the scalars are given in Table 3. Inputs are indicated by (*). The 
latin letter labels on the left hand side identify the sets in the tables.

The following empirical values are taken from \cite{RPP:2002}: 
$m_\pi^{\pm}=139.57018\pm 0.00035\ \mbox{[MeV]}$,  
$m_K^{\pm}=493.677\pm 0.016\ \mbox{[MeV]}$, 
$m_\eta=547\pm 0.12\ \mbox{[MeV]}$, 
$m_{\eta'}=957.78\pm 0.14\ \mbox{[MeV]}$ for the masses in the 
low lying pseudoscalar sector. The weak decay constants
$F^{exp}_\pi=130.7\pm 0.1\pm 0.36\ \mbox{[MeV]}$, 
$F^{exp}_K =159.8\pm 1.4\pm 0.44\ \mbox{[MeV]}$ relate to ours 
through a $\sqrt{2}$ normalization factor, thus 
$f_\pi^{exp} \simeq 92.4\ \mbox{MeV}$ and 
$f_K^{exp} \simeq 113\ \mbox{MeV}$.

The scalar masses up to $\simeq 2\ \mbox{GeV}$ are presently known to be:   
$a_0(980) =984.7\pm 1.2\ \mbox{[MeV]}$, 
$a_0(1450) =1474\pm 19\ \mbox{[MeV]}$, 
$f_0(600) =400-1200\ \mbox{[MeV]}$, 
$f_0(980) =980\pm 10\ \mbox{[MeV]}$, 
$f_0(1370) =1200-1500\ \mbox{[MeV]}$,
$f_0(1500) =1500\pm 5\ \mbox{[MeV]}$, 
$f_0(1710) =1713\pm 6\ \mbox{[MeV]}$, 
$K_0^*(1430) =1412\pm 6\ \mbox{[MeV]}$,
where the name of the particle is identified with its mass, in order not 
to clutter the notation. In \cite{Goebel:2000} there is reported the 
possibility of existence of a low lying strange scalar meson $K_0^*$. 
A broad resonance with mass $K_0^*(800) =797\pm 19\pm 43\ \mbox{[MeV]}$ 
is observed in \cite{Aitala:2002}.
\vspace{0.5cm}

\noindent {\small Table 3.
\\
The different fits for the masses of the scalar NJL nonet 
in units of MeV (except of the angle $\theta_\s$ which is given 
in degrees), as compared with a putative nonet family 
$a_0(980), K^*_0(800), f_0(600)$ and $f_0(980)$. 
The symbols of resonances stand for their masses. 
}
\\
\begin{center}
\begin{tabular}{|l|c|c|c|c|c|}
\hline
       &$a_0\ \{a_0(980)\}$
       &$K_0^*\ \{K^{*}_0(800)\}$
       &$\epsilon\ \{f_0(600)\}$
       &$\epsilon' \ \{f_0(980)\}$
       &$\theta_\s $
       \\ 
\hline
\hline
a &620   &933   &620  &1205  &35  \\ \hline
b &1215  &1164  &346  &1199  &14  \\ \hline
c &888   &976   &423  &1097  &22  \\ \hline
d &985*  &968   &249  &1017  &16  \\ \hline
e &900   &895   &224  &954   &18  \\ \hline
f &985*  &1050  &441  &1153  &20  \\ \hline
g &985*  &1150  &601  &1295  &22  \\ \hline
h &891   &954   &384  &1063  &22  \\ \hline
i &889   &1021  &489  &1252  &23  \\ \hline
j &1447  &1346  &399  &1364  &12  \\ \hline
k &907   &1087  &339  &1248  &24  \\ \hline
l &1036  &1009  &263  &1053  &15  \\ \hline
\end{tabular}
\end{center}
\vspace{0.5cm}

We start the discussion of the scalar and pseudoscalar sectors 
with the following special case shown in set (a). This pattern 
corresponds to $SU(3)$ breaking ($\hat{m}_u\ne \hat{m}_s$) 
without $U_A(1)$ breaking ($\kappa=0$) and has been considered 
in detail in Section 4 (see Eq.(\ref{mpi0}) for the 
pseudoscalars and Eq.(\ref{skappa0}) for the scalars). 

The overall description of mass spectra is reasonable, given the
simplicity of the model. Particular trends are as follows. Fixing 
$m_\pi, m_K, f_\pi,$ and $m_\eta$ (set b) or $m_{\eta'}$ (set c) to 
their empirical values, results in reducing the parameter $\kappa$ 
of the 't Hooft interaction by approximately a factor 2 in going 
from (b) to (c) (dropping slightly with increasing value of the 
cutoff). The masses for the scalars and $m_{\eta'}$ are highly 
sensitive to the choice of the $\eta$ mass: only a $4\%$ reduction 
of $m_\eta$ value in (b) is needed to get the empirical $m_{\eta'}$ 
(c), corresponding however to a $35\%$ drop of the latter with respect
to its value in (b). Fixing $\eta$ to its empirical mass in (b) not 
only yields a much too heavy $\eta'$, but also too heavy scalars 
$a_0, K_0^*$ and $\epsilon'$ (Table 3). 

Although the order of magnitude for the scalar masses in set (c) is 
reasonable, e.g. the mass of $a_0$ is obtained within $10\%$ of its 
experimental value and the $K_0^*$ mass within $20\%$, the general trend 
for a large set of parameters is $m_{a_0}<m_{K_0^*}<m_{\epsilon'}$, 
as opposed to the present empirical evidence $m_{K_0^*}<m_{a_0}\simeq 
m_{f_0(980)}$. The latter ordering can be obtained for sufficiently 
low values of $\kappa$, see set (d), with $m_{a_0}\simeq m_{\epsilon'}$ 
within $2\%$ of the empirical value, but at the expense of a very
light $\epsilon$ and too low values of current and constituent quark 
masses. The mass of $K_0^*$, being almost degenerate with $a_0$, 
remains too large by $20\%$. 

In set (e) we fix the 5 parameters of the model completely in the 
pseudoscalar sector, through $m_\pi, m_K, f_\pi, m_\eta, m_{\eta'}$. 
This constrains the $\kappa$ and $G$ parameters to comparatively 
very low values and yields also small quark masses; the $a_0$ and 
$K_0^*$ masses are almost degenerate, the $K_0^*$ mass being slightly 
smaller than the $a_0$ mass. 

In sets (f) and (g) three model parameters are fixed through $m_\pi, 
m_K, f_\pi,$ in the pseudoscalar sector and one in the scalar sector 
$m_{a_0}$, requiring that the average value of the $\eta, \eta'$ 
masses be within $10\%$ of the empirical value. 

In sets (h,i) we fix $m_\pi, m_K, f_\pi, m_{\eta'}$ and the mixing 
angle in the pseudoscalar channels. Results are also quite sensitive 
to the choice of $f_K$, see for instance sets (j) and (k), where the 
four input values of sets (b) and (c) have been kept respectively, 
fixing the remaining freedom by reducing slightly the values of $f_K$.
In set (j) a reduction of $f_K$ implies an increase in the magnitude 
of $\kappa$, increasing the splitting and turning therefore the 
$\eta'$ significantly heavier ($\eta$ remained fixed). The masses of 
the scalars increase by about $20\%$, as compared to their values in 
set (b), the lower $f_0$ a bit less, by $15\%$. In set (k) the 
reduction of $f_K$ implies also an increase in $\kappa$ and therefore 
in the splitting, this time reducing the value of $m_\eta$ (since 
$m_{\eta'}$ was kept fixed). The splitting in the scalars is also 
enhanced, the $\epsilon'$ is pushed up and $\epsilon$ down. The masses 
of $a_0$ and $K_0^*$ increase only slightly.  

In set (l) the input parameters of (b) were kept, but $f_K$ chosen 
larger. The parameter $\kappa$ gets reduced and the conclusions are
opposite to the ones of set (j).

The values of the mixing angles $\theta_\p$ and $\theta_\s$ shown in
Tables 2 and 3 are consistent with results obtained in 
\cite{Tornqvist:1999} in the framework of the linear $\sigma$ model 
with broken $U(3)\times U(3)$ symmetry, where 
$\theta_\p =-5^\circ$ and $\theta_\s =21.9^\circ$. and with the values 
$\theta_\p\approx 2^\circ$, $\phi_\s\approx -14^\circ$ reported in 
\cite{Napsuciale:2002}. The last angle here describes the mixing in 
the flavour basis and corresponds to $\bar{\psi}_\s$ (see Appendix B) 
in our notations. This agreement is not accidental, since the
bosonized NJL model is closely related to the linear sigma model
\cite{Schechter:1971,Hooft:1999}.


\section*{\normalsize 6. Concluding remarks }

We have analyzed the Nambu -- Jona-Lasinio model of QCD in the light
of a new generalized heat kernel expansion. The result is an 
effective Lagrangian of low energy QCD, incorporating the complete 
original symmetry pattern, but eliminating all non physical 
thresholds associated with quark-antiquark pair formation due to the 
lack of confinement of the original Lagrangian. We applied the so 
obtained Lagrangian in the extraction of the low lying spectra of 
pseudoscalars and scalars. The pseudoscalar spectrum turns out to 
be quite satisfactory and we used it partly to fix the main parameters 
of the model. As can be seen from Table 3 the predictions for scalar
mesons are also not too far from the experimental masses of the lightest 
known scalars, which is remarkable in view of the simplicity of the
model.  

There is growing evidence that an isovector $a_0(980)$, an isospinor 
$K^*_0(800)$, as well as two isoscalars $f_0(600)$ and $f_0(980)$, 
are members of the same low-lying scalar nonet \cite{Beveren:}, 
\cite{Pelaez:1998}, \cite{Schechter:2002}, \cite{Napsuciale:2002}. 
There are however different opinions about their origin. In our 
calculation we considered the lightest scalar nonet as being $q\bar q$
states. It is in line with ideas presented in \cite{Tornqvist:1999}.  

The outcome of the model is obtained in the leading order stationary 
phase approximation and can be implemented. There are different 
sources for corrections both at leading order and next to leading order. 
For instance the inclusion of vector and axial-vector mesons can be 
important for the physical picture, because they contribute already at 
leading order through the pseudoscalar -- axial-vector and scalar -- 
vector mixings. There are also several contributions at next to 
leading order, e.g. meson loop corrections \cite{Nikolov:1996}
and semi-classical corrections to the 't Hooft determinant 
\cite{Osipov7:2002}. As discussed in Section 4 
of \cite{Osipov7:2002} there are two distinct regimes of chiral symmetry 
breaking, related to small/large six-quark fluctuations. For large 
fluctuations the quantum corrections may be numerically relevant.  
Our aim in the present work was to show that the considered
new method for the asymptotic expansion of the heat kernel, which
is in full agreament with all symmetry requirements, leads already
in its minimal form to realistic results for mass spectra. A more 
detailed description of the scalar nonet in the framework of our 
method, including its decay properties, will be given elsewhere. 

\vspace{0.5cm}
We thank W. Broniowski for discussions and reading our manuscript.   
This work has been supported by grants provided by Funda\c c\~ao para
a Ci\^encia e a Tecnologia, POCTI/35304/FIS/2000 and 
SFRH/BPD/11579/2002.


\section*{\normalsize Appendix A. Consequences of Eq.(\ref{ha})}

The first equation in (\ref{ha}) can be written in terms of 
quark-flavour components $h_i$ (see Eq.(\ref{order-h0})). In general 
the $(u,d,s)$ basis can be transformed to the basis $(0,3,8)$ by 
the use of the following matrices $\omega_{ia}$ and $e_{ai}$ defined 
as \cite{Osipov7:2002}
\begin{equation}
\label{eo}
   e_{ai}=\frac{1}{2\sqrt 3}\left(
          \begin{array}{ccc}
          \sqrt 2&\sqrt 2&\sqrt 2 \\
          \sqrt 3&-\sqrt 3& 0 \\
          1&1&-2 
          \end{array} \right),\qquad
   \omega_{ia}=\frac{1}{\sqrt 3}\left(
          \begin{array}{ccc}
          \sqrt 2&\sqrt 3& 1 \\
          \sqrt 2&-\sqrt 3& 1 \\
          \sqrt 2&0&-2 
          \end{array} \right).
\end{equation}
Here the index $a$ runs $a=0,3,8$ (for the other values of $a$ the 
corresponding matrix elements are assumed to be zero). We have then 
for instance $h_a=e_{ai}h_i$, and $h_i=\omega_{ia}h_a$. Similar 
relations can be obtained for $\Delta_i$ and $\Delta_a$. In accordance 
with this notation we use, for instance, that 
$h^{(1)}_{ci}=\omega_{ia}h^{(1)}_{ca}$.
The following properties of matrices (\ref{eo}) are straightforward:
$\omega_{ia}e_{aj}=\delta_{ij}$, $\ e_{ai}\omega_{ib}=\delta_{ab}$, 
$e_{ai}e_{aj}=\delta_{ij}/2$ and $\omega_{ia}\omega_{ie}=2\delta_{ae}$. 
The coefficients $t_{ijk}$ are related to the coefficients $A_{abc}$ 
by the embedding formula $3\omega_{ia}A_{abc}e_{bj}e_{ck}=t_{ijk}$. 
The $SU(3)$ matrices $\lambda_a$ with index $i$ are defined in a 
slightly different way $2\lambda_i=\omega_{ia}\lambda_a$ and 
$\lambda_a=2e_{ai}\lambda_i$. In this case it follows that, for 
instance, $\sigma =\sigma_a\lambda_a=\sigma_i\lambda_i=\mbox{diag}
(\sigma_u,\sigma_d,\sigma_s)$, but $2\sigma_a\Delta_a=\sigma_i\Delta_i$.

The solutions of Eq.(\ref{order-h0}) are given in \cite{Osipov7:2002}.
One can express all other coefficients $h_{a\ldots}$ in terms of these
basic variables. We quote further our result for $h_{ab}$, splitting
the range of running indices $a,b$ on three subsets: $r,s=0,8,\  
n,m=1,2,3$ and $f,g=4,5,6,7$, 
\begin{equation}
   h_{nm}^{(1,2)}=\frac{-\delta_{nm}}{G(1\mp\omega_s)}\ , \qquad
   h_{fg}^{(1,2)}=\frac{-\delta_{fg}}{G(1\mp\omega_u)}\ . 
\end{equation}
For the $2\times 2$ matrix with indices $0,8$ we have
\begin{equation}
   h^{(1,2)}_{rs}=\frac{-1}{3G\mu_\pm}
   \left(
         \begin{array}{cc}
         3\mp (4\omega_u-\omega_s) & \
         \pm\sqrt{2}(\omega_u-\omega_s) \\
         \pm\sqrt{2}(\omega_u-\omega_s) & \
         3\pm 2(2\omega_u+\omega_s)  
         \end{array}
   \right)_{rs},
\end{equation}
with $\mu_\pm =(1\pm\omega_s-2\omega_u^2)$ and 
\begin{equation}
\label{omega}
   \omega_i=\frac{\kappa h_i}{16G}\ .
\end{equation}

Quite often the stationary phase equations considered together 
with the gap-equations help us to simplify essentially the results. 
Here is an useful example that shows Eqs.(\ref{order-h0}) and 
(\ref{gap}) at work. 

\textsf{Example.} Let us consider the expression for the mass of 
kaons following from our mesonic Lagrangian. It is not difficult 
to obtain that  
\begin{equation}
   m_K^2=g^2\left[\frac{1}{G(1+\omega_u)}+\frac{1}{2}
         \left( \frac{h_s}{m_s}+\frac{h_u}{m_u} \right)  
         \right] +(m_s-m_u)^2.
\end{equation}
One notices, by using 
\begin{equation}
   g^2\left( \frac{h_s}{m_s}-\frac{h_u}{m_u} \right)=2(m_s^2-m^2_u),
\end{equation}
which is a direct consequence of the gap equations,
that the following relation is fulfilled
\begin{equation}
   \frac{g^2}{2}
   \left( \frac{h_s}{m_s}+\frac{h_u}{m_u} \right) +(m_s-m_u)^2
   =g^2\frac{h_u+h_s}{m_u+m_s}\ .
\end{equation}  
Therefore, we obtain
\begin{equation}
   m_K^2=g^2\left( \frac{1}{G(1+\omega_u)}+\frac{h_u+h_s}{m_u+m_s}
            \right), 
\end{equation}
which can be further reduced to the final result indicated in 
the Eq.(\ref{K-mass}), by observing that 
\begin{equation}
   h_u + h_s =-\frac{\Delta_u + \Delta_s}{G(1+\omega_u)}\ .
\end{equation} 
This last expression follows immediatly from Eq.(\ref{order-h0}).


\section*{\normalsize Appendix B. Diagonalization of the mass matrix and 
                                  physical states}

To illustrate how the physical fields are chosen in the main part of 
the text we recall here some useful details of the diagonalization 
procedure and how to relate to several different conventions adopted
in the literature. Our starting point is a quadratic form ${\cal Q}$ 
written in the singlet -- octet basis $(X_0,X_8)$
\begin{equation}
\label{massmat}
   {\cal Q} = (X_0, X_8) \left( \begin{array}{cc}
                                A  & D \\
                                D  & B
                                \end{array}
                         \right)
              {X_0 \choose X_8}
\end{equation}
which can be diagonalized by an orthogonal transformation to the 
physical states $(X, \bar X)$
\begin{equation}
\label{ortog}
   {X \choose \bar X}=\left( \begin{array}{cc}
                             \cos\theta  &\sin\theta \\
                             -\sin\theta &\cos\theta
                             \end{array}
                      \right)
   {X_0 \choose X_8}.
\end{equation}
The angle $\theta$ is extracted from the equation
\begin{equation} 
\label{tg}
   \tan 2\theta =\frac{2D}{A-B}\ .
\end{equation}
After some trigonometry the $\theta$-dependence of 
the diagonalized matrix ${\cal Q}$ can be absorbed in just one term 
\begin{equation}
\label{massmat2}
   {\cal Q} = \frac{1}{2}(X, \bar X) 
              \left( \begin{array}{cc}
              A+B+\displaystyle\frac{A-B}{\cos 2\theta}  & 0 \\
              0  & A+B-\displaystyle\frac{A-B}{\cos 2\theta}
              \end{array} \right)
              {X \choose \bar X}.
\end{equation}
It is easy to see that 
\begin{equation}
   \frac{A-B}{\cos 2\theta}={\mathrm sgn}\left(
                             \frac{A-B}{\cos 2\theta}\right)
                            \sqrt{(A-B)^2+4D^2}
\end{equation} 
and therefore 
\begin{eqnarray}
\label{masses}
   &&{\cal Q}=m_X^2X^2+m_{\bar X}^2\bar{X}^2, \nonumber \\
   &&m^2_X=\frac{1}{2}\left[A+B+{\mathrm sgn}\left(
                           \frac{A-B}{\cos 2\theta}\right)
                           \sqrt{(A-B)^2+4D^2}\right], \nonumber \\
   &&m^2_{\bar X}=\frac{1}{2}\left[A+B-{\mathrm sgn}\left(
                           \frac{A-B}{\cos 2\theta}\right)
                           \sqrt{(A-B)^2+4D^2}\right].
\end{eqnarray} 
Finally, to identify the fields $X, \bar X$ with the physical ones, 
one should proceed as follows. Firstly, find the angle $\theta$ from 
Eq.(\ref{tg}), choosing, for instance, the principal value of 
$\arctan 2\theta$, i.e. $-(\pi /4)\leq \theta \leq (\pi /4)$. 
Secondly, determine the sign of the ratio $(A-B)/\cos 2\theta$.  
Only after having established which value of $(m_X, m_{\bar X})$ is 
bigger, should one proceed with identification of the physical fields, 
writing down the corresponding rotation (\ref{ortog}). 

Alternatively, one can use the nonstrange -- strange basis
$(X_\ns ,X_\s)$, where
\begin{equation}
\label{ortog2}
   {X_\ns \choose X_\s}=\frac{1}{\sqrt{3}}\left( 
                        \begin{array}{cc}
                        \sqrt{2}  & 1 \\
                        1 & -\sqrt{2}
                        \end{array}
                        \right) {X_0 \choose X_8}.
\end{equation}
Our definition (\ref{ortog}), taken together with Eq.(\ref{ortog2}), 
leads to the explicit representation 
\begin{equation}
\label{ortog3}
     {\bar X \choose X}=\left( \begin{array}{cc}
                               \cos\psi & -\sin\psi \\
                               \sin\psi &\cos\psi
                               \end{array}
                        \right)
     {X_\ns \choose X_\s},
\end{equation}
or
\begin{equation}
\label{ortog4}
     {X \choose \bar X}=\left( \begin{array}{cc}
                               \cos\bar\psi & \sin\bar\psi \\
                               -\sin\bar\psi &\cos\bar\psi
                               \end{array}
                        \right)
     {X_\ns \choose -X_\s}.
\end{equation} 
The angle $\psi$ here is equal to $\psi =\theta + 
\bar\theta_\id$, where $\bar\theta_\id$ $(\theta_\id +\bar\theta_\id 
=\pi/2)$ is determined by the equations  
$\sin\bar\theta_\id =\sqrt{2/3}$, $\cos\bar\theta_\id =1/\sqrt{3}$, 
and therefore $\psi =\theta +\arctan\sqrt{2}\simeq\theta +54.74^\circ$.
It means that $\psi$ is restricted to the range 
$9.74^\circ\lesssim\psi\lesssim 99.74^\circ$.
The angle $\bar\psi =\psi -(\pi /2)=\theta -\theta_\id$ and   
belongs to the interval
$-80.26^\circ\lesssim\bar\psi\lesssim 9.74^\circ$.
These two angles correspond to two alternative phase conventions
for a strange $\bar ss$-component. 

Here are examples that illustrate the physical interpretation of the
given formulae, using the results of our calculations obtained 
in Section 4. 

\textsf{Example 1.} In the case of pseudoscalars with broken $SU(3)$ 
symmetry but without $U_A(1)$ breaking ($\kappa =0)$ the $\phi_0$ and
$\phi_8$ components are mixed with the angle $\theta =\theta_\id$
and $(A-B)<0$. Hence, one can conclude from Eq.(\ref{masses}) that 
the $\bar X$-state is a heavier one and corresponds to $\eta'$. We 
have from (\ref{ortog3}) $\eta'\equiv -\eta_\s$ and $\eta\equiv 
\eta_\ns$. However, if $U_A(1)$ symmetry is broken $(\kappa\ne 0)$, 
one has $(A-B)>0$ (these are exactly the cases b-l shown in the Table 2) 
and we must identify the physical fields in opposite order 
$\eta'\equiv X$, $\eta\equiv\bar X$. 

\textsf{Example 2.} In the case of scalar mesons there is no 
difference between the two patterns $\kappa =0$ and $\kappa\ne 0$.
In both cases we have $({\cal A-B})<0$, i.e. $\bar X\equiv \epsilon'$,
$X\equiv \epsilon$. If $\kappa =0$, they are pure flavour states
$\bar X\equiv -\epsilon_\s$ and $X\equiv\epsilon_\ns$.



\begin{thebibliography}{99}

\bibitem{Vassilevich:2003} D.V. Vassilevich, Phys. Rept. 388 (2003) 
        270 [hep-th/0306138];
        R.D. Ball, Phys. Rept. 182 (1989) 1.
\bibitem{Ebert:1986} D. Ebert and H. Reinhardt, 
        Nucl. Phys. B271 (1986) 188.
\bibitem{Espriu:1990} D. Espriu, E. de Rafael and J. Taron,
        Nucl. Phys. B345 (1990) 22.
\bibitem{Bijnens:1993} J. Bijnens, Ch. Bruno and E. de Rafael, 
        Nucl. Phys. B390 (1993) 501.
\bibitem{Chan:1986} L.-H. Chan, Phys. Rev. Lett. 54 (1985) 1222;
        L.-H. Chan, Phys. Rev. Lett. 57 (1986) 1199;
        O. Cheyette, Phys. Rev. Lett. 55 (1985) 2394.
\bibitem{Min:1982} C. Lee, H. Min and P.Y. Pac, 
        Nucl. Phys. B202 (1982) 336;
        C. Lee, T. Lee and H. Min, Phys. Rev. D39 (1989) 1681;
        C. Lee, T. Lee and H. Min, Phys. Rev. D39 (1989) 1701.
\bibitem{Schwinger:1951} J. Schwinger, Phys. Rev. 82 (1951) 664;
        B.S. DeWitt, Dynamical Theory of Groupsand Fields, 
        Gordon and Breach, N.Y. (1965); 
        B.S. DeWitt, Phys. Rep. 19 (1975) 295.
\bibitem{Osipov1:2001} A.A. Osipov and B. Hiller, Phys. Lett. 515B 
        (2001) 458 [hep-th/0104165].
\bibitem{Osipov2:2001} A.A. Osipov and B. Hiller, Phys. Rev. D64 
        (2001) 087701  [hep-th/0106226].
\bibitem{Salcedo:2001} L.L. Salcedo, Eur. Phys. Jour. C3 (2001) 14 
        [hep-th/0107133].
\bibitem{Nambu:1961} Y. Nambu and G. Jona-Lasinio, Phys. Rev. 122 (1961)
        345; ibid 124 (1961) 246;
        V.G. Vaks and A.I. Larkin, Zh. \'{E}ksp. Teor. Fiz. 40 (1961) 
        282.  
\bibitem{Hooft:1976} A.M. Polyakov, Phys. Lett. 59B (1975) 82;
        Nucl. Phys. B120 (1977) 429; 
        A.A. Belavin, A.M. Polyakov, A. Schwartz and Y. Tyupkin,
        Phys. Lett. 59B (1975) 85;
        G. 't Hooft, Phys. Rev. Lett. 37 (1976) 8; Phys. Rev. D14 
        (1976) 3432;
        C. Callan, R. Dashen and D.J. Gross, Phys. Lett. 63B
        (1976) 334;
        R. Jackiw and C. Rebbi, Phys. Rev. Lett. 37 (1976) 172;
        S. Coleman, {\it The uses of instantons} (Erice Lectures, 1977). 
\bibitem{Veronique:1988} V. Bernard, R. L. Jaffe and U.-G. Mei\ss ner, 
        Nucl. Phys. B308 (1988) 753.
\bibitem{Reinhardt:1988} H. Reinhardt and R. Alkofer, Phys. Lett. 
        207B (1988) 482.
\bibitem{Volkov:1984} M.K. Volkov, Ann. Phys. (NY) 157 (1984) 282.
\bibitem{Osipov7:2002} A.A. Osipov and B. Hiller, Phys. Lett. 539B
        (2002) 76 [hep-ph/0204182];
        A.A. Osipov and B. Hiller, Eur. Phys. J. C 35 (2004) 223
        [hep-th/0307035].  
\bibitem{Osipov5:2001} A.A. Osipov, M. Sampaio and B. Hiller, 
        Nucl. Phys. A703 (2001) 378 [hep-ph/0110285].
\bibitem{Osipov6:2001} A.A. Osipov and B. Hiller, Phys. Rev. D63 (2001) 
        094009 [hep-ph/0012294].
\bibitem{Diakonov:1992} D. Diakonov and V. Petrov, in ``Quark Cluster
        Dynamics'', Lecture Notes in Physics, Vol. 417 (1992) 288, 
        Springer-Verlag; T. Sch\"afer and E.V. Shuryak, Rev. Mod.
        Phys. 70 (1998) 323 [hep-ph/9610451];
        D. Diakonov, [hep-ph/9602375], [hep-ph/9802298]. 
\bibitem{Carter:1999} G.W. Carter and D. Diakonov [hep-ph/9905465]. 
\bibitem{Lee:1979} C. Lee and W.A. Bardeen, Nucl. Phys. B153 (1979)
        210; M. Musakhanov, Eur. Phys. J. C9 (1999) 235 
        [hep-ph/9810295]. 
\bibitem{Diakonov:1989} D.I. Diakonov, V.Yu. Petrov and
        M. Praszalowicz, Nucl. Phys. B323 (1989) 53.
\bibitem{Gell-Mann:1968} M. Gell-Mann, R.J. Oakes and B. Renner, 
        Phys. Rev. 175 (1968) 2195.
\bibitem{Shifman:1979} M.A. Shifman, A.I. Vainshtein and V.I. Zakharov,
        Nucl. Phys. B147 (1979) 385.
\bibitem{Broniowski:1999} W. Broniowski, in ``Effective Theories of Low
        Energy QCD'', First International Workshop on Hadron Physics,
        Coimbra, Portugal, September 1999, AIP Conference Proceedings 
        508 (2000) 380 [hep-ph/9911204]. 
\bibitem{RPP:2002} Review of Particle Physics, Phys. Rev. D66 (2002).
\bibitem{Goebel:2000} C. Goebel, on behalf of the E791 Collaboration, 
        Proceedings of Heavy Quarks at Fixed Target, Rio de Janeiro, 
        October 2000, 373 [hep-ex/0012009].
\bibitem{Aitala:2002} E.M. Aitala et al [E791 Collaboration], Phys. Rev. 
        Lett. 89 (2002) 121801 [hep-ex/0204018]. 
\bibitem{Tornqvist:1999} N.A. Tornqvist, Eur. Phys. J. C11 (1999) 359 
        [hep-ph/9905282].
\bibitem{Napsuciale:2002} M. Napsuciale and S. Rodriguez, Int. J. Mod. 
        Phys. A16 (2001) 3011 [hep-ph/0204149]; 
        M. Napsuciale, A. Wirzba, M. Kirchbach, [nucl-th/0105055]; 
        M. Napsuciale, [hep-ph/0204170].
\bibitem{Schechter:1971} J. Schechter and Y. Ueda, Phys. Rev. D3 
        (1971) 2874; R. Delbourgo and M.D. Scadron, Int. J. Mod.
        Phys. A13 (1998) 657 [hep-ph/9807504]. 
\bibitem{Hooft:1999} G. 't Hooft, [hep-th/9903189].
\bibitem{Beveren:} E. van Beveren, T.A. Rjken, K. Metzger, C. Dullemond, 
        G. Rupp and J.E. Ribeiro, Z. Phys. C30 (1986) 615; 
        E. van Beveren, G. Rupp, N. Petropoulos, and F. Kleefeld,  
        Effective Theories of Low Energy QCD, 2nd Int. Workshop on Hadron 
        Physics, Coimbra, Portugal, AIP Conference Proceedings 660 (2003) 
        353.
\bibitem{Pelaez:1998} J.A. Oller, E. Oset, and J.R. Pel\'aez, Phys. ReV. 
        D59 (1999) 074001 (Erratum-ibid. D60 (1999) 099906) 
        [hep-ph/9804209]; ibid. Phys. Rev. Lett. 80 (1988) 3452 
        [hep-ph/9803242].  
\bibitem{Schechter:2002} D. Black, M. Harada and J. Schechter, Phys.
        Rev. Lett. 88 (2002) 181603 [hep-ph/0202069].
\bibitem{Nikolov:1996} N. Nikolov, W. Broniowski, C.V. Christov, 
        G. Ripka and K. Goeke, Nucl. Phys. A608 (1996) 411 
        [hep-ph/9602274];
        W. Florkowski and W. Broniowski, Phys. Lett. B386 (1996) 386
        [hep-ph/9605315].

\end{thebibliography}
\end{document}